\documentclass[aps,prl,floats,nofootinbib,twocolumn]{revtex4}
\usepackage{latexsym}
\usepackage{amssymb}
\usepackage[dvips]{graphicx}
\usepackage[applemac]{inputenc}
\usepackage{float}
\usepackage{fancyhdr}
\usepackage{amsfonts}
\usepackage{amsmath}
\usepackage{color}
\usepackage{mathrsfs}
\usepackage{bm}
\usepackage{epsfig}
\def\beq{\begin{equation}}
\def\eeq{\end{equation}}
\def\beqa{\begin{eqnarray}}
\def\eeqa{\end{eqnarray}}

\def\cm{\,{\rm cm}}

\def\s{\,{\rm s}}
\def\km{\,{\rm km}}
\def\kmps{\km\s^{-1}}

\def\pc{\,{\rm pc}}
\def\kpc{\,{\rm kpc}}

\def\kev{\,{\rm keV}}

\def\gev{\,{\rm GeV}}
\def\gevcc{\gev\cm^{-3}}

\def\half{\frac{1}{2}}

\def\rsun{\,R_0}

\def\boldx{{\bf x}}
\def\boldv{{\bf v}}
\def\boldvEarth{\bf v_{\rm E}}
\def\boldu{{\bf u}}
\def\modbx{{|\bf x|}}
\def\modbv{{|\bf v|}}
\def\modbu{{|\bf u|}}

\def\mathcalE{{\mathcal E}}
\def\mathcalF{{\mathcal F}}

\def\bv{{\bf v}}

\def\umin{u_{\rm min}}
\def\umax{u_{\rm max}}
\def\vdispsq{\langle v^2 \rangle}

\def\rhochi{\rho_{\chi}}

\def\rhodm{\rho_{\rm\scriptscriptstyle DM}}

\def\rhodmsun{\rho_{{\rm\scriptscriptstyle DM},\odot}}
\def\Phibulge{\Phi_{\rm bulge}}
\def\Phidisk{\Phi_{\rm disk}}
\def\Phidm{\Phi_{\rm\scriptscriptstyle DM}}
\def\Phivm{\Phi_{\rm\scriptscriptstyle VM}}

\def\vc{v_c}

\def\vcobsi{v^i_{\rm c,obs}}

\def\vcthi{v^i_{\rm c,th}}

\def\vcerri{v^i_{\rm c,error}}

\def\v0{\,v_0}
\def\vmax{v_{\rm max}}

\def\vmaxsun{v_{{\rm max},\odot}}

\def\vcsun{v_{c,\odot}}

\def\betam{\beta_{\rm max}}
\def\rhodm{\rho_{\rm\scriptscriptstyle DM}}

\def\Phidm{\Phi_{\rm\scriptscriptstyle DM}}
\def\Phivm{\Phi_{\rm\scriptscriptstyle VM}}

\def\rs{r_s}
\def\rhodmsun{\rho_{{\rm\scriptscriptstyle DM},\odot}}
\def\rhob0{\,\rho_{\rm\scriptscriptstyle b0}}
\def\blgsl{\,r_{\rm\scriptscriptstyle b}}
\def\dsh{\,z_{\rm\scriptscriptstyle d}}
\def\dsl{\,R_{\rm\scriptscriptstyle d}}
\def\dsd{\,\Sigma_\odot}

\def\mchi{m_\chi}
\def\mchimin{m_{\rm\scriptscriptstyle \chi,min}}
\def\msun{\,M_\odot}
\def\rvir{r_{\rm\scriptstyle vir}}

\def\fsun{f_\odot}
\def\fvsun{f_\odot (v)}

\def\Eth{E_{\rm th}}

\def\boldu{{\bf u}}

\def\mathcalR{{\mathcal R}}

\def\mathcalE{{\mathcal E}}

\def\dRdE_R{\frac{d{\mathcalR}}{dE_R}}
\def\dsigmadE_R{\frac{d\sigma}{dE_R}}
\begin{document}
\title{\bf Deriving the velocity distribution of Galactic 
Dark Matter particles\\
from rotation curve data}
\author{Pijushpani Bhattacharjee$^{1,2}$\footnote{pijush.bhattacharjee@saha.ac.in}, 
Soumini Chaudhury$^1$\footnote{soumini.chaudhury@saha.ac.in}, 
Susmita Kundu$^1$\footnote{susmita.kundu@saha.ac.in} 
and Subhabrata Majumdar$^3$\footnote{subha@tifr.res.in}}
\affiliation{
$^1$AstroParticle Physics \& Cosmology Division and Centre for AstroParticle Physics,
Saha Institute of Nuclear Physics,~1/AF Bidhannagar,~Kolkata~700064.~India\\
$^2$McDonnell center for the Space Sciences \& Department of Physics,
Washington University in St. Louis, Campus Box 1105, 
One Brookings Drive, St. Louis, MO 63130. USA\\ 
$^3$Department of Theoretical Physics, Tata Institute of Fundamental Research, Homi Bhabha 
Road, Mumbai~400005.~ India
}
\begin{abstract}
\noindent 
The velocity distribution function (VDF) of the hypothetical Weakly Interacting Massive 
Particles (WIMPs), currently the most favored candidate for the Dark Matter (DM) in the 
Galaxy, is determined directly from the circular speed (``rotation") curve data of the 
Galaxy assuming isotropic VDF. This is done by ``inverting" --- using Eddington's method 
--- the Navarro-Frenk-White universal density profile of the DM halo of the Galaxy, 
the parameters of which are determined, by using Markov Chain Monte Carlo (MCMC) technique, 
from a recently compiled set of observational data on the Galaxy's rotation curve extended 
to distances well beyond the visible edge of the disk of the Galaxy. The derived 
most-likely local isotropic VDF strongly differs from the Maxwellian 
form assumed in the ``Standard Halo Model" (SHM) customarily used in the 
analysis of the results of WIMP direct-detection experiments. 
A parametrized (non-Maxwellian) form of the derived most-likely local VDF is 
given. The astrophysical ``g-factor" that determines the effect of the WIMP VDF on the expected 
event rate in a direct-detection experiment can be lower for the derived most-likely VDF 
than that for the best Maxwellian fit to it by as much two orders of magnitude at the 
lowest WIMP mass threshold of a typical experiment. 
\end{abstract}
\maketitle
Several experiments worldwide are currently trying to directly detect the    
hypothetical Weakly Interacting Massive Particles (WIMPs), thought to 
constitute the Dark Matter (DM) halo of our Galaxy, by looking for nuclear 
recoil events due to scattering of WIMPs off nuclei of suitably chosen 
detector materials in low background underground facilities. 
The rate of nuclear recoil events depends crucially on the local (i.e., solar 
neighborhood) density and velocity distribution of the WIMPs in the 
Galaxy~\cite{Jungman_etal1996_LewinSmith_1996}, which are {\it a priori} unknown. 
Estimates based on a variety of observational data typically yield values for the local 
density of DM, $\rhodmsun$, in the range 0.2 
-- 0.4 $\gevcc$ ($(0.527 - 1.0) $ $\times 10^{-2} \msun\, \pc^{-3}$)~\cite{rhodmsun_refs}. In 
contrast, not much knowledge directly based on observational data is available on the likely 
form of the velocity distribution function (VDF) of 
the WIMPs in the Galaxy. The standard practice is to use what is often referred to as the 
``Standard Halo Model" (SHM), in which the DM halo of the Galaxy is 
described as a single-component isothermal sphere~\cite{Binney-Tremaine}, for which the 
VDF is assumed to be isotropic and of Maxwell-Boltzmann 
(hereafter simply ``Maxwellian") form, $f(\bv)\propto\exp(-{\modbv}^2/{\v0}^2)$, with a 
truncation at an assumed  value of the local escape speed, and with $\v0=\vcsun$, the 
circular rotation velocity at the location of the Sun. Apart from several theoretical issues 
(see, e.g., \cite{cbc_jcap_2010_kb_prd_2012}) concerning the self-consistency of the SHM as 
a model of a finite-size, finite-mass DM halo of the Galaxy,   
high resolution cosmological simulations of DM 
halos~\cite{VDF_sims} give strong 
indications of significant departure of the VDF from the Maxwellian. On the other hand, 
these cosmological simulations do not yet satisfactorily include the gravitational effects 
of the visible matter components of the real Galaxy, namely, the central bulge and the disk, 
which provide the dominant gravitational potential in the inner regions of the Galaxy 
including the solar neighborhood region. 

The VDF of the DM particles at any location in the Galaxy is self-consistently related to 
their spatial density as well as to the {\it total} gravitational potential, $\Phi(\boldx)$, 
at that location. For a spherical system of collisionless particles (WIMPs, for example) 
with isotropic VDF satisfying the collisionless Boltzmann equation, the Jeans 
theorem~\cite{Binney-Tremaine} ensures that the phase space distribution function (PSDF), 
$\mathcalF(\boldx,\boldv)$, depends on the phase space coordinates ($\boldx$, $\boldv$) only 
through the total energy (per unit mass), $E=\half v^2 + \Phi (r)$, where $v=\modbv$, 
$r=\modbx$. For such a system, given a isotropic spatial density distribution 
$\rho(r)\equiv\int d^3\bv \mathcalF(E)$, one can get a unique $\mathcalF$ by the Eddington 
formula~\cite{Eddington_1916,Binney-Tremaine} 
\begin{equation}
\mathcalF(\mathcalE)=\frac{1}{\sqrt{8}\pi^2}\left[\int^{\mathcalE}_0 
\frac{d\Psi}{\sqrt{\mathcalE-\Psi}}\frac{d^2\rho}{d\Psi^2}+
\frac{1}{\sqrt{\mathcalE}}\left(\frac{d\rho}{d\Psi}\right)_{\Psi=0}\right]\,,
\label{eq:Eddington_formula}
\end{equation} 
where $\Psi(r)\equiv -\Phi(r)+\Phi(r=\infty)$ is the relative potential and  
$\mathcalE\equiv -E+\Phi(r=\infty)=\Psi(r)-\frac{1}{2}v^2$ is the relative energy, with 
$\mathcalF >0$ for $\mathcalE>0$, and $\mathcalF=0$ for $\mathcalE\leq 0$. The latter  
condition implies that at any location $r$, the VDF, $f_r(\bv)=\mathcalF/\rho(r)\,,$ has a 
natural truncation at a maximum value of $v$, namely, $\vmax(r)= \sqrt{2\Psi(r)}$. 

Thus, given a isotropic density profile of a set of collisionless particles, we can 
calculate the VDF, $f_r(\bv)$, using equation (\ref{eq:Eddington_formula}) provided the 
total gravitational potential $\Phi(r)$ in which the particles move is known. A direct 
observational probe of $\Phi(r)$ is provided by the rotation curve (RC) of the Galaxy, 
the circular velocity of a test particle as a function of the galactocentric distance. In 
this paper we reconstruct the total gravitational potential $\Phi(r)$ in the Galaxy 
directly from the Galactic RC data and then use equation (\ref{eq:Eddington_formula}) to 
obtain the VDF, $f_r(\bv)$, of the  WIMPs at any location in the Galaxy~\cite{fn1}. 

We shall assume that the DM density profile to be used on the right hand side of equation 
(\ref{eq:Eddington_formula}) is of the universal NFW~\cite{NFW96} form, which, when normalized to DM density at solar location, 
$\rhodmsun$, can be written as 
\begin{eqnarray}
\rhodm(r)=\rhodmsun \left(\frac{\rsun}{r}\right)\left(\frac{\rs+\rsun}{\rs+r}\right)^2\,, 
\label{eq:rhodm_nfw_sun_norm}
\end{eqnarray}
where $\rsun$ is the distance of Sun from the Galactic centre. The profile 
(\ref{eq:rhodm_nfw_sun_norm}) has two free parameters, namely, the density $\rhodmsun$ and 
the scale radius $\rs$. 

The {\it total} gravitational potential seen by the DM particle, $\Phi$, is 
given by $\Phi=\Phidm + \Phivm$, where $\Phidm$ is the DM potential corresponding to 
the density distribution (\ref{eq:rhodm_nfw_sun_norm}) and $\Phivm$ is the total potential 
due to the visible matter (VM) component of the Galaxy. The latter can be effectively 
modeled~\cite{CO81&KG89} in terms of a spheroidal bulge superposed on an axisymmetric disk, 
with density distributions given, respectively, by 
${\rm Bulge:}\, \rho_b=\rhob0\left(1 +(r/\blgsl)^2\right)^{-3/2}\,,$
where $\rhob0$ and $\blgsl$ are the central density and scale radius of the bulge, 
respectively, and
${\rm Disk:}\, \rho_d(R,z)=\frac {\dsd}{2\dsh} e^{-(R-\rsun)/\dsl} \,\,\, 
e^{-|z|/\dsh}\,,$
where $R$ and $z$ are the axisymmetric cylindrical coordinates with $r=(R^2+z^2)^{1/2}\,$, 
$\dsl$ and $\dsh$ are the scale length and scale height of the disk, respectively, and 
$\dsd$ is its local surface density. The corresponding gravitational potentials 
for these density 
models, $\Phibulge$ and $\Phidisk$, can be easily obtained by numerically solving the 
respective Poisson equations, giving $\Phivm = \Phibulge + \Phidisk$. 

The density models specified above have a total of seven free parameters, namely, $\rs$, 
$\rhodmsun$, $\rhob0$, $\blgsl$, $\dsd$, $\dsl$, and $\dsh$. We determine the most-likely 
values and the 68\% C.L. upper and lower ranges of these parameters by performing a Markov 
Chain Monte Carlo (MCMC) analysis (see, e.g., Refs.~\cite{MCMC}) using 
the observed RC data of the Galaxy. For a given set of the Galactic model 
parameters, the circular rotation speed, $\vc(R)$, as a function of the Galactocentric 
distance $R$, is given by 
\begin{eqnarray}
\vc^2 (R) = R\frac{\partial}{\partial R}\Big[\Phidm(R,z=0)+\Phivm(R,z=0)\Big]\,. 
\label{eq:v_c_def}
\end{eqnarray}
For the observational data, we use a recently compiled set of RC  
data~\cite{Sofue_2012} that extends to Galactocentric distances well beyond the visible edge 
of the Galaxy. This data set corresponds to a choice of the Local Standard of Rest (LSR) 
set to $(\rsun,\vcsun) = (8.0\kpc, 200\kmps)$~\cite{fn2}. 
For the MCMC analysis, we use the $\chi^2$-test statistic defined as  
$\chi^2\equiv\sum_{i=1}^{i=N}\left({\frac{\vcobsi-\vcthi}{\vcerri}}\right)^2\,,$
where $\vcobsi$ and $\vcerri$ are, respectively, the observational value of the circular 
rotation speed and its error at the $i$-th value of the galactocentric distance, and 
$\vcthi$ is the corresponding theoretically calculated circular rotation speed. 
For priors on the free parameters involved, we have taken the following ranges of the 
relevant parameters based on currently available observational knowledge :  
For the VM parameters, $\rhob0:[0.1-2]\times 
4.2\times10^{2}\msun\pc^{-3}$~\cite{CO81&KG89};
$\blgsl:[0.01-0.2]\times 0.103 \kpc$~\cite{CO81&KG89};
$\dsd:[35-58]\msun\pc^{-2}$~\cite{deBoer_Weber_2010};
$\dsl:[1.7-3.5]\kpc$~\cite{Hammer_2007,CO81&KG89}. The parameter 
$\dsh$ has been fixed at $340 \pc$~\cite{surface_height_ref} since the results 
are fairly insensitive to this parameter. For the DM parameters we took a wide enough prior 
range for $\rs:[0.1-100]\kpc$ and  $\rhodmsun:[0.1-0.5]\gev\cm^{-3}$ consistent with 
values recently quoted in literature~\cite{rhodmsun_refs}. 

\begin{figure}
\centering
\epsfig{file=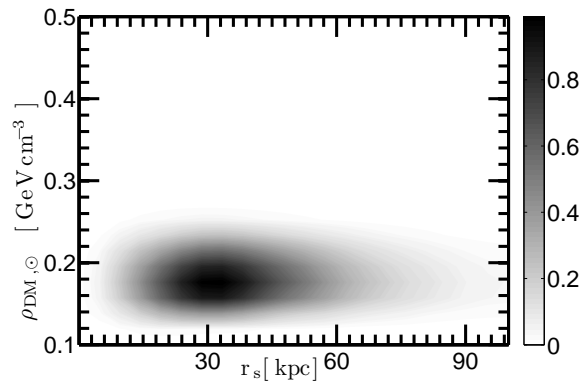,angle=0,width=3in}
\caption{The 2D posterior probability density function for Dark Matter 
parameters $(\rs - \rhodmsun)$, marginalized over the visible matter parameters.}
\label{Fig:DM2Dmcmc}
\end{figure}

\begin{figure}
\centering
\epsfig{file=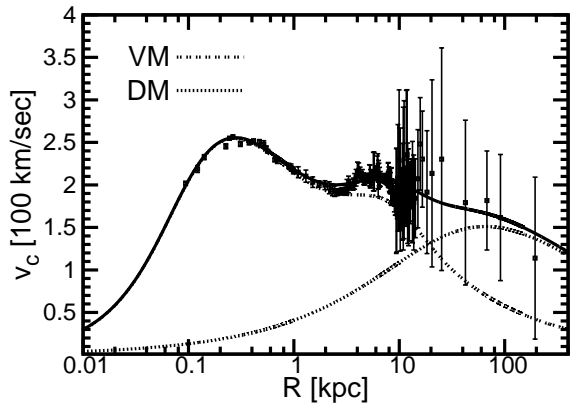,angle=270,width=3in}
\caption{Rotation curve of the Galaxy with the most-likely set of values of the 
Galactic model parameters listed in Table~\ref{Table:mcmc_in_out}. The data with 
error bars are from Ref.~\cite{Sofue_2012}.} 
\label{Fig:rotcurve}
\end{figure}

The results of our MCMC analysis are summarized in Table~\ref{Table:mcmc_in_out} and 
Figure~\ref{Fig:DM2Dmcmc}.  
\begin{table}[h]
\begin{tabular}{|c|c|c|c|c|c|c|}
\hline
 Parameter    &  $\rs$             &$\rhodmsun$     & $\rhob0\times10^{-4}$     & $\blgsl$     
& $\dsd$         & $\dsl$   \\
  Units       &  $\tiny \kpc$      & $ \gev/\cm^3$  & $\gev/\cm^3$              & $\kpc$       
& $\msun/\pc^2$  & $\kpc$   \\
\hline
Most-likely      & $30.36$            & $0.19$         & $1.83$                    & $0.092$      
& $57.9$         & $3.2$    \\
Lower         & $14.27$            & $0.17$         & $1.68$                    & $0.083$      
& $55.51$        & $2.99$    \\
Upper         & $53.37$            & $0.23$         & $2.0$                     & $0.102$      
& $58.0$         & $3.27$    \\
Mean          & $41.35$            & $0.20 $        & $1.84$                    & $0.092$      
& $54.30 $       & $3.14$    \\
SD            &  $20.51$           &  $0.02$        & $0.059$                   & $0.001$      
&  $ 3.47$       &  $0.11$   \\
\hline
\end{tabular}
\caption{The most-likely values of the Galactic model parameters, as well as 
their 68\% C.L. lower and upper ranges, means and standard deviations (SD), obtained 
from our MCMC analysis using the observed rotation curve data.}
\label{Table:mcmc_in_out}
\end{table}
Figure \ref{Fig:rotcurve} shows the theoretically calculated rotation curve for the 
most-likely 
set of values of the Galactic model parameters obtained from the MCMC analysis and listed in 
Table \ref{Table:mcmc_in_out}, and its comparison with the observed rotation curve data.  
In Table~\ref{Table:bfit_model_mass_with_range}, we display the 
values of some of the physical quantities of interest characterizing the Galaxy, derived 
from the Galactic parameters listed in Table~\ref{Table:mcmc_in_out}. 
\begin{table}[h]
\begin{tabular}{|l|c|c|}
\hline
Derived Quantities  & Unit & Values \\
\hline
\scriptsize Bulge mass ($M_b$) 				  & $10^{10}\msun$  & 
$3.53^{+1.81}_{-1.29}$\\
\scriptsize Disk mass ($M_d$)  				  & $10^{10}\msun$  & 
$4.55^{+0.2}_{-0.22}$\\
\scriptsize Total VM mass ($M_{\rm VM}=M_b+M_d$) 	  & $10^{10}\msun$  & 
$8.07^{+2.01}_{-1.51}$\\
\scriptsize DM Halo virial radius ($\rvir$) 		  & $\kpc$ 	    & 
$199.0^{+75}_{-53.5}$\\
\scriptsize Concentration parameter ($\frac{\rvir}{\rs}$) & $-$ 	    & 
$6.55^{+5.01}_{-2.05}$\\
\scriptsize DM halo virial mass ($M_h$) 	  	  & $10^{11}\msun$  & 
$8.61^{+14.01}_{-5.22}$ \\
\scriptsize Total mass of Galaxy ($M_{\rm VM}+M_h$)	  & $10^{11}\msun$  & 
$9.42^{+14.21}_{-5.37}$\\
\scriptsize DM mass within $\rsun$		    	  & $10^{10}\msun$  & 
$1.89^{+0.72}_{-0.3}$\\
\scriptsize Total mass within $\rsun$		  	  & $10^{10}\msun$  & 
$7.09^{+1.9}_{-1.15}$\\
\scriptsize Total surface density :			  &		    &  			   
\\
\scriptsize at $\rsun$ ($|z|\leq 1.1\kpc$)		  & $\msun\pc^{-2}$ & 
$69.21^{+2.52}_{-3.55}$\\
\scriptsize Total Mass within $60 \kpc$   	          & $10^{11}\msun$  & 
$3.93^{+2.15}_{-1.41}$\\
\scriptsize Total Mass within $100 \kpc$  	          & $10^{11}\msun$  & 
$5.92^{+4.35}_{-2.56}$\\
\scriptsize Local Circular velocity ($\vcsun$)	          & $\kmps$ 	    & 
$206.47^{+24.67}_{-16.3}$\\
\scriptsize Local maximum velocity ($\vmaxsun$)           & $\kmps$	    & 
$516.02^{+120.85}_{-97.58}$\\
\hline
\end{tabular}
\caption{The most-likely values of various relevant physical 
parameters of the Milky Way and their upper and lower ranges 
derived from the most-likely- and 68\% C.L. upper and lower ranges of values of the Galactic 
model parameters listed in Table~\ref{Table:mcmc_in_out}.}
\label{Table:bfit_model_mass_with_range}
\end{table}
The values in Table~\ref{Table:bfit_model_mass_with_range} are in reasonably good 
agreement with the values of these quantities quoted in recent 
literature~\cite{Catena_Ullio_2012,Sofue_2012,TableII_numbers_refs}.
The relatively large uncertainties in the values of some of the quantities that receive 
dominant contribution from the DM halo properties at large Galactocentric distances are 
simply a reflection of the relatively large uncertainties of the rotation curve data at 
those distances.  
%

The Galactic model parameters determined above allow us to reconstruct the total 
gravitational potential $\Phi(\boldx)$ at any location in the Galaxy. Because of the 
axisymmetric nature of the VM disk, this potential is non-spherical. To use equation 
(\ref{eq:Eddington_formula}), which is valid only for a spherical symmetric situation, we  
use the spherical approximation~\cite{Ullio_Kamion_2001,Catena_Ullio_2012}, 
$\Phivm(r)\simeq G\int_0^r M_{\rm VM}(r')/r'^2 dr'$, where $M_{\rm VM}$ is the total VM 
mass contained within $r$ \cite{fn3}. 

\begin{figure}
\centering
\epsfig{file=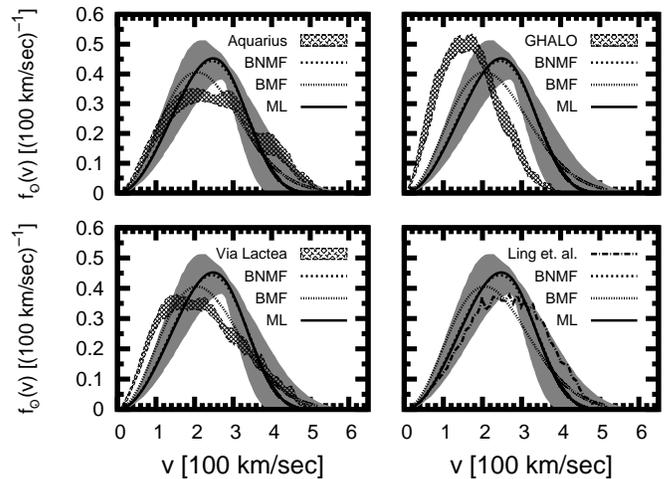,angle=270,width=3.8in}
\caption{Normalized local speed distribution, $\fvsun$, corresponding to the most-likely 
(ML) set of values of the Galactic model parameters given in Table \ref{Table:mcmc_in_out} 
(solid curve) and its uncertainty band (shaded) corresponding to the 68\% 
C.L.~upper and lower ranges of the Galactic model parameters. The four panels 
show comparison of our results with those from four different N-body 
simulations~\cite{VDF_sims}, as indicated. In each panel, the best non-Maxwellian 
fit (BNMF: equation \ref{eq.prop_model} --- almost indistinguishable from the ML curve) as 
well as the best Maxwellian fit (BMF), the latter with the form $f_\odot^{\rm 
Maxwell}(v)\propto v^2 \exp\left(-v^2/\v0^2\right)$ truncated at $\vmaxsun=516\kmps$ (see 
Table \ref{Table:bfit_model_mass_with_range}) and with the free parameter $\v0$ determined 
to be $206\kmps$, are also shown.} 
\label{Fig:fv_fit_simu_ana}
\end{figure}

The resulting normalized speed distribution, $f_r(v)\equiv \left(4\pi 
v^2\right)f_r(\bv)$ (with $\int f_r(v)dv = 1$), 
evaluated at the location of the Sun, giving the most-likely $\fvsun$, is shown in 
Figure \ref{Fig:fv_fit_simu_ana}. For comparison, we also show in the same Figure the 
best Maxwellian fit (with $f_\odot^{\rm Maxwell}(v)\propto 
v^2\exp\left(-v^2/\v0^2\right)$) to the most-likely $\fvsun$ obtained from MCMC analysis.   
We also compare our results with those from four large N-body simulations~\cite{VDF_sims}. 

As evident from Figure \ref{Fig:fv_fit_simu_ana}, the speed distribution differs 
significantly from the Maxwellian form. We find that the following parametrized 
form, which goes over to the standard Maxwellian form in the limit of the parameter $k\to 
0$, gives a good fit to our numerically obtained most-likely local speed distribution shown 
in Figure \ref{Fig:fv_fit_simu_ana}:  
\beq
\fvsun \approx  4\pi v^2 \left(\xi(\beta)-\xi(\betam)\right)\,,
\label{eq.prop_model}
\eeq
where $\xi(x)=(1+x)^k~e^{-x^{(1-k)}}$, $\beta=v^2/\v0^2$, $\betam=\vmaxsun^2/\v0^2$, 
$\v0=339\kmps$ and $k=-1.47$. As a quantitative measure of the 
deviation of a model form of the local speed distribution, $f^{\rm model}$, from the 
numerically obtained most-likely (ML) form, $f^{\rm ML}$, shown in Figure 
\ref{Fig:fv_fit_simu_ana}, the quantity $\chi^2_f \equiv (1/N)\sum_{i=1}^N
\left[f^{\rm ML}(v_i) - f^{\rm model}(v_i)\right]^2$ 
has a value of $\sim 7.2\times10^{-5}$ for the parametrized form (\ref{eq.prop_model}) 
compared to a value $\sim1.7\times10^{-3}$ for the best Maxwellian fit shown in Figure 
\ref{Fig:fv_fit_simu_ana}. Note also that our results differ significantly from those 
obtained from the N-body simulations. 

\begin{figure}
\centering
\epsfig{file=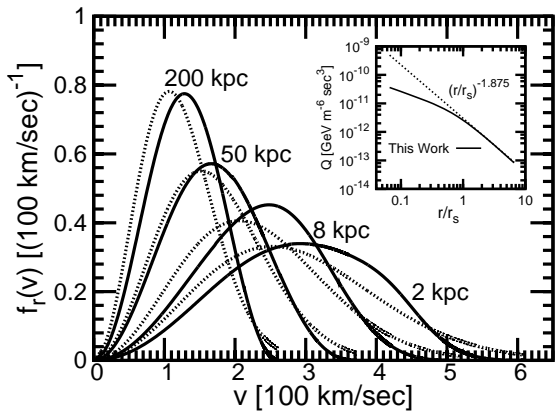,angle=270,width=3in}
\caption{Normalized speed distribution of the DM particles at various Galactocentric 
radii (solid curves), corresponding to the most-likely set of values of the Galactic model 
parameters given in Table \ref{Table:mcmc_in_out}. The curves (dotted) for the corresponding 
best Maxwellian fit are also shown for comparison. The inset shows the 
pseudo-phase space density of DM, $Q\equiv\rho/\vdispsq^{3/2}$, as 
a function of $r$.}
\label{Fig:fv_diff_r_ppsd}
\end{figure}

In Figure \ref{Fig:fv_diff_r_ppsd} we show the most-likely $f_r(v)$'s at 
several different values of the Galactocentric distance $r$. Notice how the peak of the 
distribution shifts towards smaller values of $v$ and the width of the distribution shrinks,  
as we go to larger $r$, with the distribution eventually becoming a delta function at zero 
speed at asymptotically large distances, as expected. The non-Maxwellian nature of the 
distribution at all locations is also clearly seen, with the Maxwellian approximation always 
overestimating the number of particles at both low as well as extreme high 
velocities. The inset in Figure \ref{Fig:fv_diff_r_ppsd} shows our results for the 
pseudo phase space density, $Q\equiv\rho/\vdispsq^{3/2}$, as a function of 
$r$, and its comparison with the power-law behavior predicted from simulation 
results~\cite{ppsd_refs}. Note the agreement with the power-law behavior at large distances 
but strong deviation from it at smaller Galactocentric radii, which we attribute to the 
effect of the visible matter: For a given DM density profile, the additional gravitational 
potential provided by the VM supports higher velocity dispersion of the DM 
particles, making $Q$ smaller than that for the DM-only case. 

We now discuss the implications of our results for the analysis of direct detection 
experiments. The differential rate of nuclear recoil events per unit detector
mass (typically measured in counts/day/kg/keV), in which a
WIMP (hereafter generically denoted by $\chi$ with mass $\mchi$) elastically scatters off a
target nucleus of mass $m_N$ leaving the recoiling nucleus with a kinetic energy
$E_R$, can be written as~\cite{Jungman_etal1996_LewinSmith_1996}
\beq
\frac{d \mathcalR}{dE_R}(E_R,t) = \frac{\sigma(q^2=2m_NE_R)}{2\mchi\mu^2}\rhochi
g(E_R,t)\,,
\label{eq:recoil_rate_def}
\eeq
where $\rhochi\equiv\rhodmsun$ is the local mass density of WIMPs, 
$\sigma(q^2)$ is the momentum transfer dependent effective WIMP-nucleus elastic cross section, 
$\mu=\mchi m_N/(\mchi + m_N)$ is the reduced mass of the WIMP-nucleus
system, and
\beq
g(E_R,t)= \int_{u>\umin(E_R)}^{\umax(t)} \frac{d^3\boldu}{u} 
\fsun\left(\boldu+\boldvEarth(t)\right)\Theta(\umax-\umin)\,,
\label{eq:g_def}
\eeq
is the crucial ``g-factor" that contains all information about the local VDF of the 
WIMPs~\cite{gmin_refs}. 
In (\ref{eq:g_def}) the variable $\boldu$ (with 
$u=\modbu$) represents the relative 
velocity of the WIMP with respect to the detector at rest on Earth, and $\boldvEarth(t)$ is 
the (time-dependent) velocity of the Earth relative to the Galactic rest frame. The quantity 
$\umin(E_R)=\left(m_N E_R/2\mu^2\right)^{1/2}$ is the minimum WIMP speed required for giving 
a recoil energy $E_R$ to the nucleus, and $\umax (t)$ is the (time-dependent) maximum WIMP 
speed~\cite{cbc_jcap_2010_kb_prd_2012} corresponding to the maximum speed $\vmax$ (defined 
in the Galactic rest frame) for the VDF under consideration. 
Note that the quantity $g(E_R,t)$ takes its largest value at $E_R=\Eth$, the threshold 
energy for the experiment under consideration. 
\begin{figure}[h]
\centering
\epsfig{file=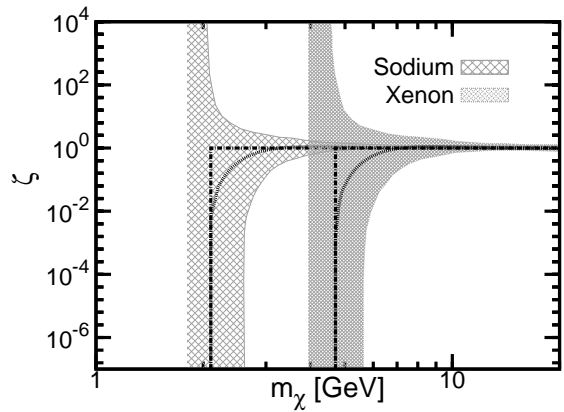,angle=270,width=3in}
\caption{The ratio (solid curves), $\zeta\equiv g_{\rm ML}(\Eth)/g_{\rm Maxwell}(\Eth)$, of 
the g-factor calculated with our most-likely (ML) form of $\fvsun$ shown in Figure
\ref{Fig:fv_fit_simu_ana} to that for the best Maxwellian fit to it also shown in Figure
\ref{Fig:fv_fit_simu_ana}, as a function of the WIMP mass $\mchi$, for two different target 
nuclei, namely, Sodium and Xenon, both with $\Eth=2\kev$. The shaded bands correspond to the 
uncertainty bands of $\fvsun$ shown in Figure \ref{Fig:fv_fit_simu_ana}. The calculations 
are for 2nd June, when the Earth's velocity in the Galactic rest frame is maximum.}  
\label{Fig:g_diff_targets}
\end{figure}

To illustrate the effect of the non-Maxwellian nature of the VDF and its 
uncertainty, we define the quantity $\zeta\equiv g_{\rm 
ML}(\Eth)/g_{\rm Maxwell}(\Eth)$, the 
ratio of the g-factor calculated with our most-likely (ML) form of $\fvsun$ shown in Figure 
\ref{Fig:fv_fit_simu_ana} to that for the best Maxwellian fit to it also shown in 
Figure 
\ref{Fig:fv_fit_simu_ana}, both evaluated at $E_R=\Eth$. A plot of $\zeta$ as a function of 
the WIMP mass $\mchi$, for two different target nuclei, viz. Sodium and Xenon, in both 
case with $\Eth=2\kev$, is shown in Figure \ref{Fig:g_diff_targets}. 

The lowest WIMP mass that can be probed by a given experiment is given 
by~$\mchimin=m_N\left[\left(2m_N (\vmaxsun + v_{\rm 
E})^2 /\Eth\right)^{1/2}-1\right]^{-1}$. As seen from Figure \ref{Fig:g_diff_targets}, the 
effect of the departure from Maxwellian distribution is most significant at the lowest WIMP 
mass where the difference can be as much as two orders of magnitude. 

To summarize, a first attempt has been made to derive the velocity distribution (assumed 
isotropic) of the dark matter particles in the Galaxy directly using the  
rotation curve data. The distribution is found to be significantly non-Maxwellian in 
nature, the implication of which is a sizable deviation of the expected direct detection 
event rates from those calculated with the usual Maxwellian form. 

{\bf Acknowledgments:} We thank Satej Khedekar and Subir Sarkar for useful 
discussions. One of us (P.B.) thanks Ramanath Cowsik for discussions and for support under 
the Clark Way Harrison Visiting Professorship program at the McDonnell Center for the 
Space Sciences at Washington University in St. Louis.  


\begin{thebibliography}{99}
\bibitem{Jungman_etal1996_LewinSmith_1996} G.~Jungman, M.~Kamionkowski and K.~Griest, 
Phys.~Rep. {\bf 267} (1996) 195; J.D.~Lewin and R.F.~Smith, Astropart.~Phys.
{\bf 6} (1996) 87.
\bibitem{rhodmsun_refs} J.H. Oort, Bull.~Astr.~Inst.~Netherlands {\bf 6} (1932) 249; {\it 
ibid.} {\bf 15} (1960) 45; J.N.~Bahcall, ApJ {\bf 276} (1984) 169; P.~Salucci, 
F.~Nesti, G.~Gentile and C.F.~Martins, A\&A {\bf 523} (2010) A83; R. Catena and P. Ullio, 
JCAP {\bf 08} (2010) 004; J.~Bovy and S.~Tremaine, ApJ {\bf 756} (2012) 89. 
 \bibitem{Binney-Tremaine} J. Binney and S. Tremaine, {\it Galactic Dynamics} (2nd Ed.), 
(Princeton University Press, Princeton, 2008). 
\bibitem{cbc_jcap_2010_kb_prd_2012} S.~Chaudhury, P. Bhattacharjee and R.~Cowsik, JCAP {\bf 
09} (2010) 020; S.~Kundu and P.~Bhattacharjee, Phys.~Rev.~D {\bf 85} (2012) 123533.  
\bibitem{VDF_sims} J.~Stadel et al., MNRAS {\bf 398} (2009) L21 (GHALO); J.~Diemand et al., 
Nature {\bf 454} (2008) 735 (Via Lactea); V.~Springel et al., MNRAS {\bf 391} (2008) 1685 
(Aquarius); F.S.~Ling et al., JCAP {\bf 1002} (2010) 012. 
\bibitem{Eddington_1916} A.S.~Eddington, MNRAS {\bf 76} (1916) 572.
\bibitem{fn1} An earlier work by Catena and Ullio~\cite{Catena_Ullio_2012} used Eddington's 
method to derive the {\it local} VDF from various dynamical constraints on 
the gross properties of the Galaxy rather than the full RC data as done here.  
\bibitem{Catena_Ullio_2012} R.~Catena and P.~Ullio, JCAP {\bf 1205} (2012) 005. 
\bibitem{NFW96} J.F.~Navarro, C.S.~Frenk and S.D.M.~White, ApJ {\bf 462} 
(1996) 563. 
\bibitem{CO81&KG89} J.~Caldwell and J.~Ostriker, ApJ {\bf 251} (1981) 61; 
K.~Kuijken and G.~Gilmore, MNRAS {\bf 239} (1989) 571; 
ibid.~{\bf 239} (1989) 605; ibid.~{\bf 239} (1989) 651; ApJ {\bf L9} (1991) 367. 
\bibitem{MCMC} A.~Putze et al., A\&A {\bf 497} (2009) 991; A.~Heavens, arXiv:0906.0664v3; 
http://cosmologist.info/cosmomc/. 
\bibitem{Sofue_2012} Y.~Sofue, PASJ {\bf 64} (2012) 75 [arXiv:1110.4431]. 
\bibitem{fn2} The exercise done in this paper can be repeated with other choices for the 
LSR. However, we do not expect the qualitative nature of our results 
to change significantly. 
\bibitem{deBoer_Weber_2010} M.~Weber and W.~de Boer, A\&A {\bf 509} (2010) A25. 
\bibitem{Hammer_2007} F.~Hammer et al., ApJ {\bf 662} (2007) 322. 
\bibitem{surface_height_ref} H.T.~Freudenreich, ApJ {\bf 492} (1998) 495. 
\bibitem{TableII_numbers_refs} P.J.~McMillan, MNRAS {\bf 414} (2011) 
2446; M.I.~Wilkinson and N.W.~Evans, MNRAS {\bf 310} (1999) 645; W.~Dehnen and J.~Binney, 
MNRAS {\bf 294} (1998) 429.  
\bibitem{Ullio_Kamion_2001} P. Ullio and M.~Kamionkowski, JHEP {\bf 03} (2001) 049.  
\bibitem{fn3} We use the numerically calculated full axisymmetric VM potential in our 
MCMC analysis to find the Galactic model parameters. From this, we estimate that the 
error in the potential due to the spherical approximation at the solar location is $\sim$ 
9\%, and the approximation gets increasingly better at larger $r$. The resulting errors 
in the normalized speed distribution gives curves which still lie within the uncertainty 
bands (due to uncertainties of the rotation curve data) shown in Figure 
\ref{Fig:fv_fit_simu_ana}.    
\bibitem{ppsd_refs} J.E.~Taylor and J.F.~ Navarro, ApJ {\bf 563} (2001) 483.
\bibitem{gmin_refs} M.~Lisanti, L.E.~Strigari, J.G.~Wacker and R.H.~Wechsler, 
Phys.~Rev~{\bf D83} (2011) 023519; P.J.~Fox, G.D.~Kribs and T.M.~Tait, Phys.~Rev~{\bf D83} 
(2011) 034007; M.T.~Frandsen et al., JCAP {\bf 01} (2012) 024. 
\end{thebibliography}
\end{document}